# DIRA : A Framework of Data Integration Using Data Quality

Reham I. Abdel Monem[1], Ali H. El-Bastawissy[2] and Mohamed M. Elwakil[3]

[1]Information Systems Department,
Faculty of computers and information, Cairo University, Giza, Egypt
[2]Faculty of Computer Science, MSA University, Giza, Egypt
[3]The Software Engineering Laboratory, Innopolis University, Innopolis, Russia

*ABSTRACT*

*Data integration is the process of collecting data from different data sources and providing user with unified view of answers that meet his requirements. The quality of query answers can be improved by identifying the quality of data sources according to some quality measures and retrieving data from only significant ones. Query answers that returned from significant data sources can be ranked according to quality requirements that specified in user query and proposed queries types to return only top-k query answers. In this paper, Data integration framework called Data integration to return ranked alternatives (DIRA) will be introduced depending on data quality assessment module that will use data sources quality to choose the significant ones and ranking algorithm to return top-k query answers according to different queries types.*

*KEYWORDS*

*Data integration, data sources, query answers, quality measures, top-k query answers, assessment module, ranking algorithm*

## 1. INTRODUCTION

Data integration is the process of combining data from multiple and heterogonous data sources in unified view to satisfy users' queries. It has different architectures but virtual integration and data warehousing architectures are the most commonly used [1]. Data warehouse is a single integrated physical source of data for processing information and it loads data through (ETL) extract, transform and load process, Virtual data integration is the process of combining several local data sources to form single virtual data source. In virtual data integration, data stores in local data sources and accesses through global schema.

Data sources have different levels of quality that specify their fitness for using in specific task, these quality levels change over time and can be measured through some data quality measures that have different classifications [2] and following we will present one of their classifications in table 1.

Table 1.Illustrates data integration IQ measures classification.[3]

| Data Integration Components | IQ Criteria |
|---|---|
| Data Source | Reputation, Verifiability, Availability and Response Time. |
| Schema | Schema Completeness, Minimalism and Type Consistency. |
| Data | Data Completeness, Timeliness, Accuracy and Data Validity. |

Data quality measures can be assessed at different granularities. First, collection of data sources level which assesses the aggregate quality for collection of data sources. Second, data source level which assesses the quality for the whole source. Third, relation level which assesses the quality for data source relations fourth, attribute level which assesses the quality for a relation attributes and we will use all these levels in our framework. They have relationships between them, these relationships are critical for effective knowledge discovery and finding these relationships or dependencies is dependency discovery. For example, the valid values must be complete values but complete values can be valid values or not.[2]

Following we will focus on data quality measures that could affect the data integration process, could be considered important from user's prospective and we will refer to them as data quality features.

**1.1 Data Completeness**

Data completeness classified in literature into two types: Null-Completeness and Population Completeness. Null-Completeness is "the degree of missing data or knowing of null values for some data". Population-Completeness is "the availability of all needed data by user" and can be classified into two types of relational model named Closed World Assumption (CWA) and Open World Assumption (OWA). In our work, we will use Population-Completeness under OWA and we will introduce a new type of completeness called Fact-Completeness.

Following, we will introduce the way to measure each type of completeness at attribute level:

Null-Completeness Assessment ( $C_{Null}$): it is the ratio between the number of non-null values (Complete values) and the total number of values or the complement value to ratio between the number of null values (InComplete values) and the total number of values. [4]

$$C_{Null} = \frac{\text{Number of non-null values}}{\text{total number of values}} \quad (1)$$

Or

$$C_{Null} = 1 - \frac{\text{Number of null values}}{\text{total number of values}} \quad (2)$$

- **Scaled Aggregate Data Completeness Value for Queried Attributes ( C )**

$$\text{Scaled Total } (C) = \frac{\sum_1^m C_{Null}(a_m)}{M} \quad (3)$$

Where M is total number of queried attributes

- Population-Completeness Assessment ($C_{Population}$): It is the ratio of tuples actually represented in a relation r, with respect to the whole number of tuples in ref(r) where ref(r), is the relation containing all tuples that satisfy the relational schema of r.[4]

$$C_{Population}(r) = \frac{Cardinality\ of\ r}{Cardinality\ of\ ref(r)} \quad (4)$$

## 1.2 Data Validity

**Data validity** is "the degree to which attribute value follows specified domain, data item isn't valid if its value is out of the domain" and can be measured at attribute level in our framework.

Data Validity Assessment ($P_{Q_{Validity}}$): It is the ratio between the number of valid values and the total number of values.[4]

$$P_{Q_{Validity}} = \frac{Number\ of\ valid\ values}{total\ number\ of\ values} \quad (5)$$

- Scaled Aggregate Data Validity Value for Queried Attributes (L)

$$\text{Scaled Total }(L) = \frac{\sum_1^m L(a_m)}{M} \quad (6)$$

Where M is total number of queried attributes

## 1.3 Data Accuracy

**Data accuracy** classified in literature into two types: semantic accuracy and (0 or 1) accuracy. Semantic accuracy refers to the degree of closeness between value v (recorded value) near to value v' (correct value), (0 or 1) accuracy will consider data values are accurate if they don't conflict with real-world values and inaccurate otherwise. In our work, we will use (0 or 1) accuracy and it will be measured at attribute level.

Data Accuracy Assessment ($P_{Q_{Accurate}}$): It is the ratio between the number of accurate values and the total number of values.[4]

$$P_{Q_{Accurate}} = \frac{Number\ of\ accurate\ values}{total\ number\ of\ values} \quad (7)$$

- Scaled Aggregate Data Accuracy Value for Queried Attributes (A)

$$\text{Scaled Total }(A) = \frac{\sum_1^m A(a_m)}{M} \quad (8)$$

Where M is total number of queried attributes

## 1.4 Data Timeliness

**Data timeliness** is " the degree to which data is up-to-date". So, it captures the gap from data creation to data delivery and can be measured at attribute level in our framework.

International Journal of Data Mining & Knowledge Management Process (IJDKP) Vol.6, No.2, March 2016

- Data Timeliness Assessment: Timeliness is assessed and rescaled according to below equations.[4]

$$\text{Currency} = \text{Age} + (\text{DeliveryTime} - \text{InputTime}) \qquad (9)$$

Currency: The degree to which data value reflects all changes that happen to it.
Age: How old the data is when it is received.
DeliveryTime: The time when data is delivered to user.
InputTime: The time when data is obtained.

$$\text{Timeliness} = \max\left\{0, 1 - \frac{Currency}{Volatility}\right\} \qquad (10)$$

Volatility: The length of time that data remains valid.

In our work, we will suppose that DeliveryTime = InputTime (no delay from obtaining data to deliver it to user) so Currency = Age

- Aggregate Data Timeliness Value for Queried Attributes (T)

$$Total(T) = \text{Maximum}(T(a_m)) \qquad (11)$$

This paper is organized as follows; section 2 will include different approaches concerned with data integration in terms of data quality, the proposed framework for data integration will be explained in section 3. The conclusion and future work will be presented in section 4.

## 2. RELATED WORK

Many approaches are developed to introduce data integration in terms of data quality. Following, we will present an overview of some approaches related to our framework; how they measure and store data quality, how they process queries and user interference option.

### 2.1 DaQuinCIS Approach

This approach designed to deal with cooperative information systems and to exchange not only intended data but also metadata. The query processing approach implemented by DaQuinCIS to return a query answer is structured as following:[4]:

1. **Query unfolding**: Each concept in user query Q that sends in terms of global schema is defined in terms of local schemas to retrieve all data that answers user query from all available participating data sources. So, Q will decompose into $Q_1,…..Qn$ queries to send to each relevant local data source to return results $R_1,….Rn$.

2. **Extensional checking**: In this step $R_1$ ∪ $R_2$….. ∪ Rn are passed to record matching algorithm to discover the same objects. The output of this step is clusters of similar objects.

3. **Result building**: In this step the best quality object representative will be chosen according to quality value q associated with each field value f. If an object contains the

40

highest quality values for all fields, so it will be chosen as representative otherwise a representative object will be constructed from combination of highest qualified fields' values within cluster. Once all representatives are chosen, the final result will be built from union of all these representatives.

This approach depends on data sources metadata to improve query answers through improving fusion process.

## 2.2 Data Quality based Data Integration Approach

This approach explains the importance of data quality in data integration. It adds quality system components to integrate data quality dimensions (completeness, accuracy, cost and response time) to data integration system for selecting less number of single data sources for more qualified query results.

This approach presents experiments using Amalgam and THALIA benchmarks to show that the query results delivered in a reasonable amount of time and using the minimum number of possible data sources.[1]

The concept of this approach will be used but to retrieve highest top-k qualified query results from significant data sources only according to different proposed queries types.

## 3. DATA INTEGRATION TO RETURN RANKED ALTERNATIVES (DIRA) FRAMEWORK

In this section, we will illustrate our new data integration framework called DIRA. DIRA data quality assessment module will be presented in section 3.1, new completeness type (Fact-Completeness) will be explained in section 3.2, DIRA quality system components will be introduced in section 3.3 and DIRA workflow components will be explained in detail in section 3.4.

## 3.1 DIRA Data Quality Assessment Module

This module consists of different components to evaluate data quality, these components are[5]:

- **Assessment Metrics** are procedures for calculating data quality features and estimates an assessment score for these features using a scoring function.

- **Aggregation Metrics** are procedures for calculating aggregated score from distinct assessment scores using aggregation functions like sum, count, and average functions.

- **Data Quality Features** are meta-data for providing user with indication of how data fit to task at hand.

- **Scoring Functions** are the way for calculating data quality features. There are many scoring functions to choose between them like simple comparisons, set function, aggregation function and complex statistical function.

### 3.2 Data Fact-Completeness

Data fact-completeness is a new accurate type of completeness where it uses null-completeness and population-completeness to assess its value.

- Fact-Completeness Assessment ($C_{fact}$): It is subtraction of the probability of incomplete values from the probability of population-completeness values. We will present its equation on attribute level and then we will aggregate the values for higher levels.

$$C_{Population}(a_m \ (r)) = \frac{Cardinality\ of\ a_m\ (r)}{Cardinality\ of\ ref(r)} \quad (12)$$

$$InC_{Null}(a_m \ (r)) = \frac{Count\ of\ null\ values\ for\ a_m}{Cardinality\ of\ ref(r)} \quad (13)$$

$$C_{fact}(a_m \ (r)) = C_{Population}\ (a_m \ (r)) - InC_{Null}\ (a_m \ (r)) \quad (14)$$

Where $a_m$ (r) refers to attribute number m in relation r

- **Scaled Aggregate Data Fact-Completeness Value for Queried Attributes ( C )**

$$\text{Scaled Total } (C) = \frac{\sum_1^m C_{fact}(a_m)}{M} \quad (15)$$

Where M is total number of queried attributes

### 3.3 DIRA Quality System Components

In our work, we will add some components to integration systems called quality system components to improve query answers. These components are data quality acquisition and user input.

#### 3.3.1 Data Quality Acquisition

This component is responsible for storing attributes and relations found in data sources in metadata store. It is also responsible for running data quality queries and storing their answers in the metadata store

DIRA Metadata Store that is presented in figure1 will use the concept of hierarchical quality framework [6] to build its entities that we will explain in table 2.

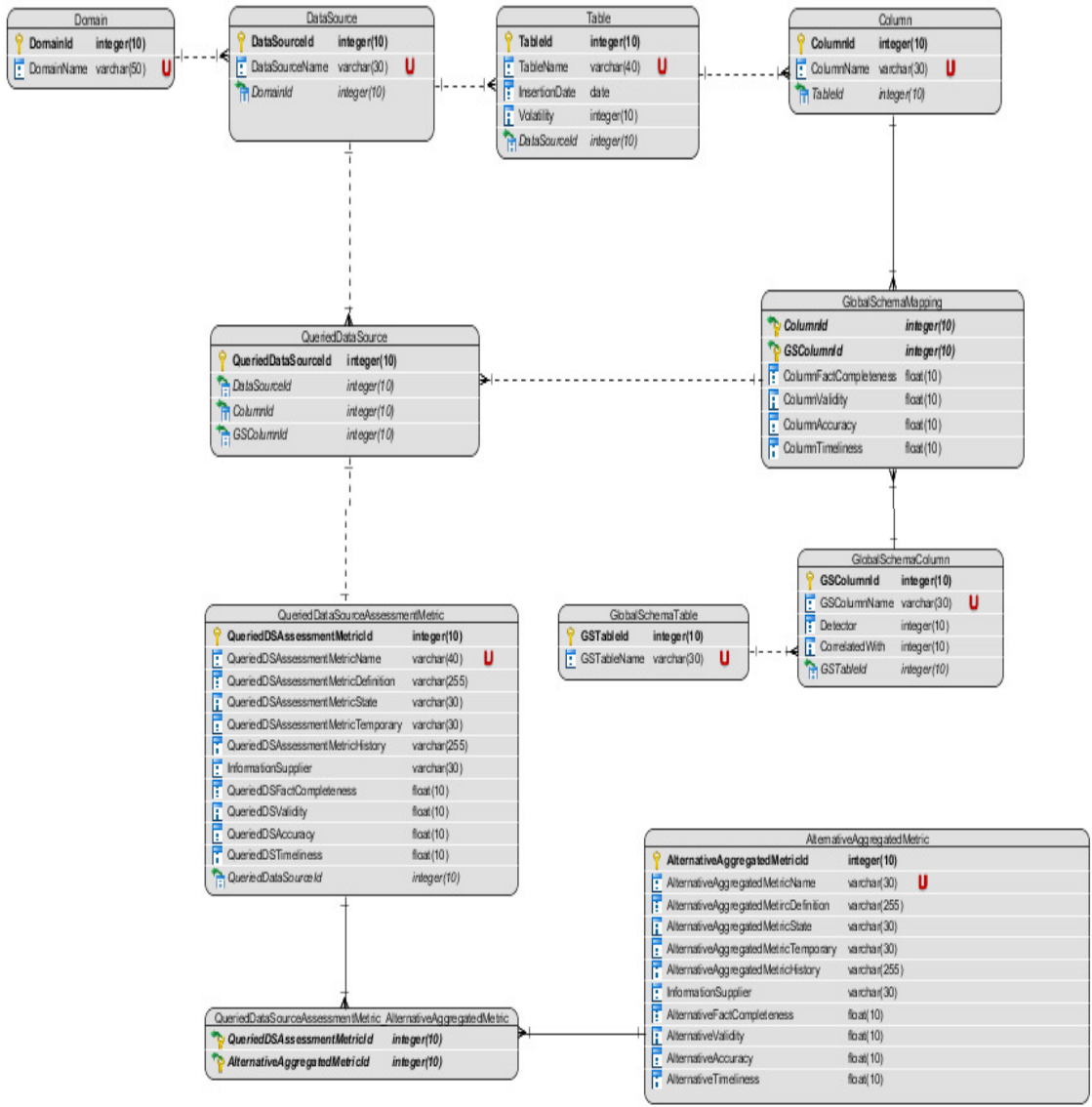

Figure 1.DIRA metadata structure

Table 2.Presents DIRA metadata structure entities

| Entity | Definition |
| --- | --- |
| DataSource | This entity contains information about all data sources participating in data integration process. |
| Domain | This entity contains information about data sources' domains. |
| Table | This entity contains information about all data sources' tables. |
| Column | This entity contains information about all tables' columns. |
| GlobalSchemaTable | This entity contains information about all tables in global schema. |
| GlobalSchemaColumn | This entity contains information about all columns in every table in global schema. |
| GlobalSchemaMapping | - This entity is associative entity<br>- It contains information about tables' columns with their correspondence in global schema.<br>- It contains scores that evaluate the data quality for every data source column with its correspondence in global schema according to scoped data quality features, these scores will be calculated once during data integration system configuration, they will be updated according to data sources modification and they will be used in evaluating the results that will return from queried data sources without returning data for early pruning to these data sources (data integration will retrieve data from only sources that can answer query and can satisfy the required level of quality).<br>-Scores assessment will save time and cost for data integration process especially for data sources with high volatility. |
| QueriedDataSource | This entity contains information about data sources that can participate with attributes in query answering and which attributes it can participate with. |
| QueriedDataSourceAssessmentMetric | This entity contains information about every data source that can participate in query answering and its data quality features' total scores (this entity represents evaluation for data that every data source can participate with in answering query). |
| AlternativeAggregatedMetric | -This entity contains information about qualified alternatives (qualified alternative is one or more queried data source that can answer query and can satisfy the required level of quality if specified) for given query.<br>-It contains qualified alternatives data quality features scores (aggregated scores for alternatives with more than one data source and assessment scores for alternatives with one data source).<br> -Qualified alternatives will pass to ranking algorithm to return top-k ranking alternatives before duplicate detection and data fusion. |
| QueriedDataSourceAssessmentMetric_ AlternativeAggregatedMetric | This entity is associative entity that contains the IDs of qualified alternatives aggregated metrics and IDs of their related queried data sources assessment metrics. |

### 3.3.2 User Input

SQL can be extended to include some quality constrains that will be required by user in query to return qualified results, these constrains are expressed by data quality features. Query Q syntax with quality constraint [1]

```
Select A1... Ak
From G
Where < selection condition >
With < data quality goal >
Where A1.A2, Ai are global attributes of G
```

## 3.4 DIRA Workflow Components

In this section, we will explain in detail the DIRA workflow components (Data Sources Attributes (columns) Assessment Metrics, Queried Data Sources Assessment Metrics, Alternative Formation, Alternatives Aggregated Metrics and Alternatives Ranking) and they will be presented in figure2.

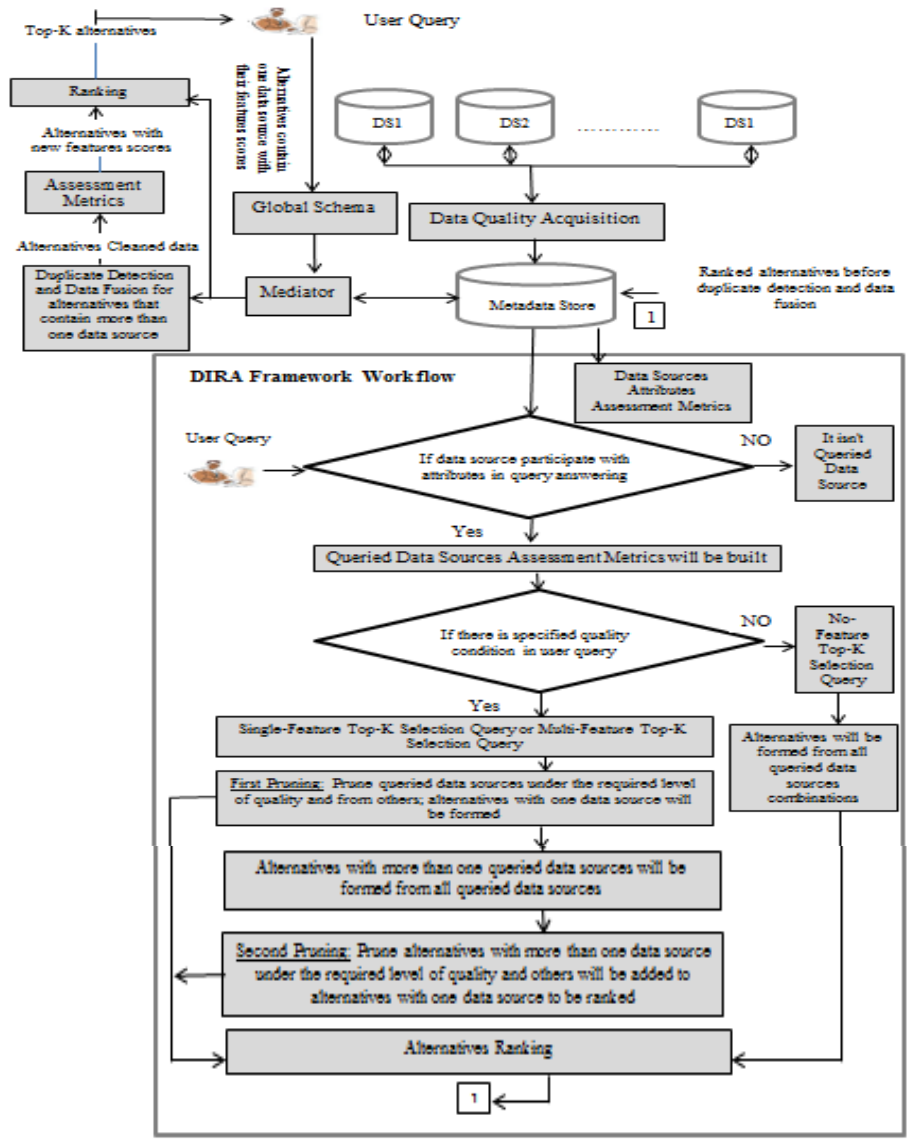

Figure 2. DIRA workflow components

DIRA components will be explained in the following motivation example1. Example1 represents three data sources DS1, DS2 and DS3 with their data and the status of these data from data completeness, data validity and data accuracy (Note: The assessment date was on 2/2/2016). In relation data; one refers to value status and two refers to the consequences of this status.

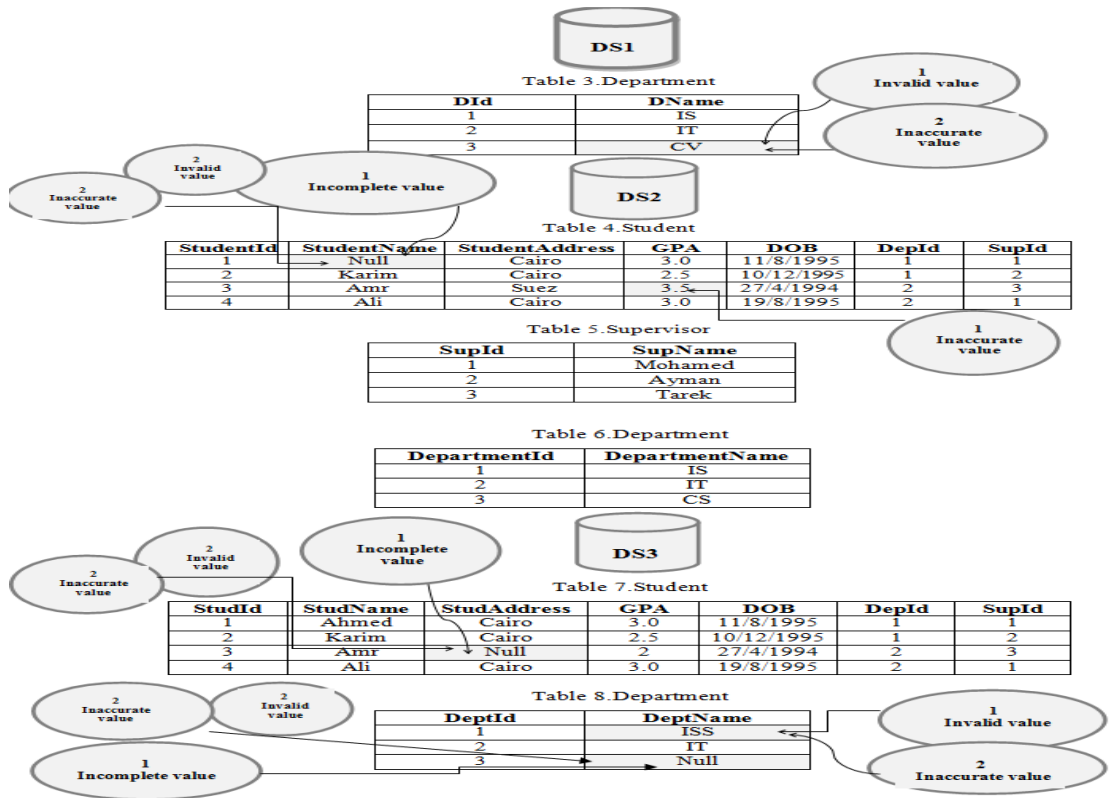

Suppose that the **reference relations** for data sources relations are

Table 9.Student

| SupId | SName | SAddress | GPA | DOB | DepId | SupId |
|---|---|---|---|---|---|---|
| 1 | Ahmed | Cairo | 3.0 | 11/8/1995 | 1 | 1 |
| 2 | Karim | Cairo | 2.5 | 10/12/1995 | 1 | 2 |
| 3 | Amr | Suez | 2.0 | 27/4/1994 | 2 | 3 |
| 4 | Ali | Cairo | 3.0 | 19/8/1995 | 2 | 1 |

Table 10.Supervisor

| SupId | SupName |
|---|---|
| 1 | Mohamed |
| 2 | Ayman |
| 3 | Tarek |

Table 11.Department

| DepartmentId | DepartmentName |
|---|---|
| 1 | IS |
| 2 | IT |
| 3 | CS |

### 3.4.1 Data Sources Attributes Assessment Metrics

In this component, we will assess the scoped data quality features scores for all attributes (columns) in data sources with its correspondence in global schema as we presented in global schema mapping entity. Following are tables that we will use to build this component

Table 12.Domain entity

| DomainID | DomainName |
|---|---|
| 1 | Cairo University |

Table 13.DataSource entity

| DataSourceId | DataSourceName | DomainID |
|---|---|---|
| 1 | DS1 | 1 |
| 2 | DS2 | 1 |
| 3 | DS3 | 1 |

Table 14.Table entity

| TableId | TableName | InsertionDate | Volatility | DataSourceId |
|---|---|---|---|---|
| 1 | Department | 2/12/2015 | 365 | 1 |
| 2 | Student | 2/1/2016 | 365 | 2 |
| 3 | Supervisor | 2/1/2016 | 365 | 2 |
| 4 | Department | 2/1/2016 | 365 | 2 |
| 5 | Student | 2/10/2015 | 365 | 3 |
| 6 | Department | 2/10/2015 | 365 | 3 |

Table 15.Column entity

| ColumnId | ColumnName | TableId |
|---|---|---|
| 1 | DId | 1 |
| 2 | DName | 1 |
| 3 | StudentId | 2 |
| 4 | StudentName | 2 |
| 5 | StudentAddress | 2 |
| 6 | GPA | 2 |
| 7 | DOB | 2 |
| 8 | SupId | 3 |
| 9 | SupName | 3 |
| 10 | DepartmentId | 4 |
| 11 | DepartmentName | 4 |
| 12 | StudId | 5 |
| 13 | StudName | 5 |
| 14 | StudAddress | 5 |
| 15 | GPA | 5 |
| 16 | DOB | 5 |
| 17 | DeptId | 6 |
| 18 | DeptName | 6 |

Table 16.GlobalSchemaTable entity

| GSTableId | GSTableName |
|---|---|
| 1 | Student |
| 2 | Supervisor |
| 3 | Department |

Table 17.GlobalSchemaColumn entity

| GSColumnId | GSColumnName | Detector | CorelatedWith | GSTableId |
|---|---|---|---|---|
| 1 | SId | 0 | Null | 1 |
| 2 | SName | 1 | Null | 1 |
| 3 | SAddress | 1 | Null | 1 |
| 4 | GPA | 0 | Null | 1 |
| 5 | DOB | 1 | Null | 1 |
| 6 | SupId | 0 | Null | 2 |
| 7 | SupName | 0 | Null | 2 |
| 8 | DId | 0 | Null | 3 |
| 9 | DName | 0 | Null | 3 |

According to equations 5, 7, 9, 10, 12, 13 and 14 that we presented in section 1 and section 3.2, the GlobalSchemaMapping entity will be presented in table 18.

**N.P:** According to data quality features' relationships[2], we can deduce the following relation (1) to validate values in table 18.

$$P_{Q_{Completeness}} \geq P_{Q_{Validity}} \geq P_{Q_{Accuracy}} \qquad (1)$$

Table 18.GlobalSchemaMapping entity

| Column Id | GSColumn Id | Column Population Completness | Column InCompletness | Column FactCompletness | Column Validity | Column Accuracy | Column Timeliness |
|---|---|---|---|---|---|---|---|
| 3 | 1 | 1 | 0 | 1 | 1 | 1 | 0.92 |
| 12 | 1 | 1 | 0 | 1 | 1 | 1 | 0.67 |
| 4 | 2 | 1 | 0.25 | 0.75 | 0.75 | 0.75 | 0.92 |
| 13 | 2 | 1 | 0 | 1 | 1 | 1 | 0.67 |
| 5 | 3 | 1 | 0 | 1 | 1 | 1 | 0.92 |
| 14 | 3 | 1 | 0.25 | 0.75 | 0.75 | 0.75 | 0.67 |
| 6 | 4 | 1 | 0 | 1 | 1 | 0.75 | 0.92 |
| 15 | 4 | 1 | 0 | 1 | 1 | 1 | 0.67 |
| 7 | 5 | 1 | 0 | 1 | 1 | 1 | 0.92 |
| 16 | 5 | 1 | 0 | 1 | 1 | 1 | 0.67 |
| 8 | 6 | 1 | 0 | 1 | 1 | 1 | 0.92 |
| 9 | 7 | 1 | 0 | 1 | 1 | 1 | 0.92 |
| 1 | 8 | 1 | 0 | 1 | 1 | 1 | 0.84 |
| 10 | 8 | 1 | 0 | 1 | 1 | 1 | 0.92 |
| 17 | 8 | 1 | 0 | 1 | 1 | 1 | 0.67 |
| 2 | 9 | 1 | 0 | 1 | 0.67 | 0.67 | 0.84 |
| 11 | 9 | 1 | 0 | 1 | 1 | 1 | 0.92 |
| 18 | 9 | 1 | 0.33 | 0.67 | 0.33 | 0.33 | 0.67 |

**3.4.2 Queried Data Sources Assessment Metrics**

In this component first, we will fill **QueriedDataSource** associative entity table 19 with data related to each query.

Table 19.QueriedDataSource

| QueriedDataSourceId | DataSourceId | ColumnId | GSColumnId |
|---|---|---|---|
| 1 | 2 | 4 | 2 |
| 2 | 3 | 13 | 2 |
| 3 | 2 | 5 | 3 |
| 4 | 3 | 14 | 3 |
| 5 | 2 | 7 | 5 |
| 6 | 3 | 16 | 5 |
| 7 | 2 | 9 | 7 |
| 8 | 1 | 2 | 9 |
| 9 | 2 | 11 | 9 |
| 10 | 3 | 18 | 9 |

Q1: Select SName, SAddress, DOB, SupName, DName From G

Second, we will fill **QueriedDataSourceAssessmentMetric** entity table 20 with data.

International Journal of Data Mining & Knowledge Management Process (IJDKP) Vol.6, No.2, March 2016

Table 20. Illustrative table

| Queried DataSource Id | Queried Column Id | Queried GSColumn Id | Queried GSColumn Name | Queried Column FactCompleteness | Queried Column Validity | Queried Column Accuracy | Queried Column Timeliness |
|---|---|---|---|---|---|---|---|
| 1 | 2 | 9 | DName | 1 | 0.67 | 0.67 | 0.84 |
| 2 | 4 | 2 | SName | 0.75 | 0.75 | 0.75 | 0.92 |
| 2 | 5 | 3 | SAddress | 1 | 1 | 1 | 0.92 |
| 2 | 7 | 5 | DOB | 1 | 1 | 1 | 0.92 |
| 2 | 9 | 7 | SupName | 1 | 1 | 1 | 0.92 |
| 2 | 11 | 9 | DName | 1 | 1 | 1 | 0.92 |
| 3 | 13 | 2 | SName | 1 | 1 | 1 | 0.67 |
| 3 | 14 | 3 | SAddress | 0.75 | 0.75 | 0.75 | 0.67 |
| 3 | 16 | 5 | DOB | 1 | 1 | 1 | 0.67 |
| 3 | 18 | 9 | DName | 0.67 | 0.33 | 0.33 | 0.67 |

According to equations 6, 8, 11 and 15 that we presented in section 1 and section 3.2, the QueriedDataSourceAssessmentMetric entity for Q1 will be presented in table 21

Table 21. QueriedDataSourceAssessmentMetric entity

| QueriedDS Assessment MetricId | Queried Data Source Id | QueridDS Assessment MetricName | QueriedDS Assessment Metric Definition | QueriedDS Assessment Metric Temporary | QueriedDS Assessment MetricState | QueriedDS Assessment Metric History |
|---|---|---|---|---|---|---|
| 1 | 1 | DS1Assessment Metric | This metric is evaluation for DS1 required attributes data quality | Dynamic | Positive | This evaluation metric is for DS1 that found in Queried DataSource entity with its required attributes |
| 2 | 2 | DS2Assessment Metric | This metric is evaluation for DS2 required attributes data quality | Dynamic | Positive | This evaluation metric is for DS2 that found in Queried DataSource entity with its required attributes |
| 3 | 3 | DS3Assessment Metric | This metric is evaluation for DS3 required attributes data quality | Dynamic | Positive | This evaluation metric is for DS3 that found in Queried DataSource entity with its required attributes. |

| QueriedDS Assessment Metric Id | Information Supplier | QueriedDS Fact Completeness | QueriedDS Validity | QueriedDS Accuracy | QueriedDS Timeliness |
|---|---|---|---|---|---|
| 1 | User | 0.20 | 0.13 | 0.13 | 0.84 |
| 2 | User | 0.95 | 0.95 | 0.95 | 0.92 |
| 3 | User | 0.68 | 0.62 | 0.62 | 0.67 |

### 3.4.3 Alternatives Formation

Users became not interested in how to access different data sources or how to combine the results from them. Users' requested data can be found in single source, in different sources or distributed between many sources.

**Alternative** represents one or more queried data source that participates in data integration, it may be qualified alternative or not qualified alternative.
**N/P:** In our framework,

- Alternatives formation will be specified according to query type.

- For queries without any required quality condition, alternatives will be formed from combinations of all queried data sources and all will be considered as qualified alternatives.
- For queries with quality condition, we will consider every queried data source that satisfies the required quality level as qualified alternative of single queried data source and we will prune others from forming alternatives from one queried data source (First Pruning).We will build combinations from all queried data sources to form alternatives from two or three or more queried data sources, alternatives that will not satisfy the required level of quality will be pruned (Second Pruning) and the remains will be considered as qualified alternatives.
- Total number of alternatives before first and second pruning will be within $\{0,\ldots\ldots\ldots 2^M - 1\}$.

**Alternatives formation according to Q2 (Q2 with quality condition)**

```
Q2: Select SName, SAddress, DOB, SupName, DName
From G
With AlternativeFactCompleteness ≥ 0.65
Order by AlternativeFactCompleteness desc
Limit 3
```

**First pruning according to Q2:** from QueriedDataSourceAssessmentMetric table 21, we can specify that DS1 will prune from forming alternative alone (not qualified alternative) as presented in table 23 because it is under the level of quality specified in user query Q2.

**Qualified alternatives with one queried data source: Alternative2:** (DS2), **Alternative3:** (DS3)

Table 22.QueriedDataSourceAssessmentMetric_AlternativeAggregatedMetric entity for Q2 after first pruning

| QueriedDSAssessmentMetricId | AlternativeAggregatedMetricId |
|---|---|
| 2 | 2 |
| 3 | 3 |

Table 23.AlternativeAggregatedMetric entity for Q2 after first pruning

| Alternative Aggregated MetricId | Alternative Aggregated MetricName | Alternative Aggregated MetricDefinition | Alternative Aggregated MetricTemporary | Alternative Aggregated MetricState | Alternative Aggregated Metric History | Information Supplier |
|---|---|---|---|---|---|---|
| 2 | Alternative2 | This alternative contains DS2Assessment Metric and it can satisfy the required level of data quality where query answer completeness≥0.65 | Dynamic | Positive | This alternative evaluated by Ds2 Assessment Metric that found in QueriedData Source Assessment Metric entity | User |
| 3 | Alternative3 | This alternative contains DS3Assessment Metric and it can satisfy the required level of data quality where query answer completeness≥0.65 | Dynamic | Positive | This alternative evaluated by Ds3 Assessment Metric that found in QueriedData Source Assessment Metric entity | User |

| Alternative Aggregated MetricId | Alternative Fact Completeness | Alternative Validity | Alternative Accuracy | Alternative Timeliness |
|---|---|---|---|---|
| 2 | 0.95 | 0.95 | 0.95 | 0.92 |
| 3 | 0.68 | 0.62 | 0.62 | 0.67 |

International Journal of Data Mining & Knowledge Management Process (IJDKP) Vol.6, No.2, March 2016

**Second pruning according to Q2:** from QueriedDataSourceAssessmentMetric table 21, we will build combinations from queried data sources to form alternatives from two or three or more queried data sources as we will introduce in table 24 and we will prune alternatives that will not satisfy the required level of quality according to their calculated aggregated metrics that will presented in section 3.4.4.

**Alternatives from two or more queried data sources: Alternative4**: (DS1, DS2), **Alternative5**: (DS1, DS3), **Alternative6**: (DS2, DS3) and **Alternative7**: (DS1, DS2, DS3)

Table 24.QueriedDataSourceAssessmentMetric_AlternativeAggregatedMetric entity for Q2 before second punning

| QueriedDSAssessmentMetricId | AlternativeAggregatedMetricId |
|---|---|
| 2 | 2 |
| 3 | 3 |
| 1 | 4 |
| 2 | 4 |
| 1 | 5 |
| 3 | 5 |
| 2 | 6 |
| 3 | 6 |
| 1 | 7 |
| 2 | 7 |
| 3 | 7 |

### 3.4.4 Alternatives Aggregated Metrics

In this component, aggregated scores will be calculated for alternatives with two or more queried data sources from their assessment metrics.

Following, we will present equations to assess our scoped data quality features aggregated scores for alternatives where Q is number of data sources that form the alternative[7]

- The fact completeness of alternative (collections of DBs)

$$C_{DBs} = \sum_{q=1}^{Q} C_{fact}(DS_q)/Q \qquad (16)$$

- The data validity of alternative (collections of DBs)

$$L_{DBs} = \sum_{q=1}^{Q} L(DS_q)/Q \qquad (17)$$

- The data accuracy of alternative (collections of DBs)

$$A_{DBs} = \sum_{q=1}^{Q} A(DS_q)/Q \qquad (18)$$

- The data timeliness of alternative (collections of DBs)

$$TotalAlternativeT = \text{Maximum}(T(DS_q) \text{ in Alternartive}) \qquad (19)$$

By applying equations 16, 17, 18 and 19 on Queried Data Sources Assessment Metrics table 21 according to Q2, The Alternatives Aggregated Metrics for Alternative4, Alternative5, Alternative6 and Alternative7 will be presented in table 25

International Journal of Data Mining & Knowledge Management Process (IJDKP) Vol.6, No.2, March 2016

Table 25. AlternativeAggregatedMetric entity for Q2 before second pruning

| Alternative Aggregated MetricId | Alternative Aggregated MetricName | Alternative Aggregated MetricDefinition | Alternative Aggregated MetricTemporary | Alternative Aggregated MetricState | Alternative Aggregated Metric History | Information Supplier |
|---|---|---|---|---|---|---|
| 2 | Alternative2 | This alternative contains DS2Assessment Metric and it can satisfy the required level of data quality where query answer completeness≥0.65 | Dynamic | Positive | This alternative evaluated by Ds2 Assessment Metric that found in QueriedData Source Assessment Metric entity | User |
| 3 | Alternative3 | This alternative contains DS3Assessment Metric and it can satisfy the required level of data quality where query answer completeness≥0.65 | Dynamic | Positive | This alternative evaluated by Ds3 Assessment Metric that found in QueriedData Source Assessment Metric entity | User |
| 4 | Alternative4 | This alternative contains DS1 and DS2 Assessment Metrics | Dynamic | Positive | This alternative can be evaluated by DS1 and DS2 Assessment Metrics that found in QueriedData Source Assessment Metric entity | User |
| 5 | Alternative5 | This alternative contains DS1 and DS3 Assessment Metrics | Dynamic | Positive | This alternative can be evaluated by DS1 and DS3 Assessment Metrics that found in QueriedData Source Assessment Metric entity | User |
| 6 | Alternative6 | This alternative contains DS2 and DS3 Assessment Metrics | Dynamic | Positive | This alternative can be evaluated by DS2 and DS3 Assessment Metrics that found in QueriedData Source Assessment Metric entity | User |
| 7 | Alternative7 | This alternative contains DS1, DS2 and DS3 Assessment Metrics | Dynamic | Positive | This alternative can be evaluated by DS1, DS2 and DS3 Assessment Metrics that found in QueriedData Source Assessment Metric entity | User |

| Alternative Aggregated MetricId | Alternative Fact Completeness | Alternative Validity | Alternative Accuracy | Alternative Timeliness |
|---|---|---|---|---|
| 2 | 0.95 | 0.95 | 0.95 | 0.92 |
| 3 | 0.68 | 0.62 | 0.62 | 0.67 |
| 4 | 0.58 | 0.54 | 0.54 | 0.92 |
| 5 | 0.44 | 0.38 | 0.38 | 0.54 |
| 6 | 0.82 | 0.79 | 0.79 | 0.92 |
| 7 | 0.61 | 0.54 | 0.57 | 0.92 |

According to the required quality level in Q2, we will prune Alternative4, Alternative5 and Alternative7 as they are under the required level of quality and the final qualified alternatives metrics for Q2 will be presented in table 26.

Table 26. AlternativeAggregatedMetric entity for Q2 after second pruning

| Alternative Aggregated MetricId | Alternative Aggregated MetricName | Alternative Aggregated MetricDefinition | Alternative Aggregated MetricTemporary | Alternative Aggregated MetricState | Alternative Aggregated Metric History | Information Supplier |
|---|---|---|---|---|---|---|
| 2 | Alternative2 | This alternative contains DS2Assessment Metric and it can satisfy the required level of data quality where query answer completeness≥0.65 | Dynamic | Positive | This alternative evaluated by Ds2 Assessment Metric that found in QueriedData Source Assessment Metric entity | User |
| 3 | Alternative3 | This alternative contains DS3Assessment Metric and it can satisfy the required level of data quality where query answer completeness≥0.65 | Dynamic | Positive | This alternative evaluated by Ds3 Assessment Metric that found in QueriedData Source Assessment Metric entity | User |
| 6 | Alternative6 | This alternative contains DS2 and DS3Assessment Metrics and it can satisfy the required level of data quality where query answer completeness≥0.65 | Dynamic | Positive | This alternative can be evaluated by DS2 and Ds3 Assessment Metric that found in QueriedData Source Assessment Metric entity | User |

| Alternative Aggregated MetricId | Alternative Fact Completeness | Alternative Validity | Alternative Accuracy | Alternative Timeliness |
|---|---|---|---|---|
| 2 | 0.95 | 0.95 | 0.95 | 0.92 |
| 3 | 0.68 | 0.62 | 0.62 | 0.67 |
| 6 | 0.82 | 0.79 | 0.79 | 0.92 |

51

International Journal of Data Mining & Knowledge Management Process (IJDKP) Vol.6, No.2, March 2016

### 3.4.5 Alternatives Ranking

For many years, the advantages of databases and information retrieval systems have merged to achieve the goal of many researchers. While database systems provide efficient treatment with data, mechanisms for effective retrieval and fuzzy ranking[8] that are more attractive to the user are provided with IR. In our work, we will rank alternatives according to their data quality features scores and according to different queries types

### 3.4.5.1 Ranking Alternatives according to Proposed Queries Types

We will present different queries' types depending on number of quality features in query condition (from one to four) and the kind of quality features' value (quantitative or qualitative). Quantitative data are values presented as numbers and qualitative data are values presented by a name, symbol, or a number code and they require user intervention as low, medium and high as features' values can represent different scores to different users.

### I. No-Feature Top-K Selection Query

In this type, queries don't include any specified quality condition, so we will build AlternativeAggregatedMetric table 28 from QueriedDataSourceAssessmentMetric table 27. Then, we will return to user all alternatives ranked according to all proposed features as presented in table 29, table 30, table 31, and table 32 and we will let him to choose the most suitable feature ranking

**User Query (Q3)**  
Q3: Select SName, SAddress, DOB, SupName  
From G  
Limit 3

Table 27. QueriedDataSourceAssessmentMetric entity for Q3

| QueriedDS Assessment MetricId | Queried Data Source Id | QueridDS Assessment MetricName | QueriedDS Assessment MetricDefinition | QueriedDS Assessment Metric Temporary | QueriedDS Assessment MetricState | QueriedDS Assessment Metric History |
|---|---|---|---|---|---|---|
| 1 | 1 | DS2Assessment Metric | This metric is evaluation for DS2 required attributes data quality | Dynamic | Positive | This evaluation metric is for DS2 that found in QueriedDataSource entity with its required attributes |
| 2 | 2 | DS3Assessment Metric | This metric is evaluation for DS3 required attributes data quality | Dynamic | Positive | This evaluation metric is for DS3 that found in QueriedDataSource entity with its required attributes. |

| QueriedDS Assessment Metric Id | Information Supplier | QueriedDS FactCompleteness | QueriedDS Validity | QueriedDS Accuracy | QueriedDS Timeliness |
|---|---|---|---|---|---|
| 1 | User | 0.95 | 0.95 | 0.95 | 0.92 |
| 2 | User | 0.68 | 0.62 | 0.62 | 0.67 |

International Journal of Data Mining & Knowledge Management Process (IJDKP) Vol.6, No.2, March 2016

Table 28. AlternativeAggregatedMetric entity for Q3

| AlternativeAggregatedMetricId | AlternativeAggregatedMetricName | AlternativeAggregatedMetricDefinition | AlternativeAggregatedMetricTemporary | AlternativeAggregatedMetricState | AlternativeAggregatedMetricHistory | InformationSupplier |
|---|---|---|---|---|---|---|
| 1 | Alternative1 | This alternative contains DS2AssessmentMetric | Dynamic | Positive | This alternative can be evaluated by Ds2 Assessment Metric that found in QueriedDataSource AssessmentMetric entity | User |
| 2 | Alternative2 | This alternative contains DS3AssessmentMetric | Dynamic | Positive | This alternative can be evaluated by Ds3 Assessment Metric that found in QueriedDataSource AssessmentMetric entity | User |
| 3 | Alternative3 | This alternative contains D2 and DS3AssessmentMetrics | Dynamic | Positive | This alternative can be evaluated by DS2 and Ds3 Assessment Metric that found in QueriedData Source AssessmentMetric entity | User |

| AlternativeFactCompleteness | AlternativeValidity | AlternativeAccuracy | AlternativeTimeliness |
|---|---|---|---|
| 0.95 | 0.95 | 0.95 | 0.92 |
| 0.68 | 0.62 | 0.62 | 0.67 |
| 0.82 | 0.79 | 0.79 | 0.92 |

Table 29. Top 3 Alternatives Ranked according to FactCompleteness

| AlternativeName | AlternativeFactCompleteness |
|---|---|
| Alternative1 | 0.95 |
| Alternative3 | 0.82 |
| Alternative2 | 0.68 |

Table 30. Top 3 Alternatives Ranked according to Validity

| AlternativeName | AlternativeValidity |
|---|---|
| Alternative1 | 0.95 |
| Alternative3 | 0.79 |
| Alternative2 | 0.62 |

Table 31. Top 3 Alternatives Ranked according to Accuracy

| AlternativeName | AlternativeAccuracy |
|---|---|
| Alternative1 | 0.95 |
| Alternative3 | 0.79 |
| Alternative2 | 0.62 |

Table 32. Top 3 Alternatives Ranked according to Timeliness

| AlternativeName | AlternativeTimeliness |
|---|---|
| Alternative1 | 0.92 |
| Alternative3 | 0.92 |
| Alternative2 | 0.67 |

**I. Single-Feature Top-K Selection Query**

In this type, queries include one data quality feature as a quality condition, so we will build AlternativeAggregatedMetric table from QueriedDataSourceAssessmentMetrice table but after first and second pruning according to specified quality condition in user query and then return to user alternatives ranked according to required data quality feature in user query.

This type of queries classified into two categories Quantitate Single-Feature Top-K Selection Query and Qualified Single-Feature Top-K Selection Query. We will introduce every category as following:

- Quantitate Single-Feature Top-K Selection Query

In Quantitate Single-Feature Top-K Selection Query, Data quality features values are presented as quantitate values.

A SQL template for Single-Feature top-k selection query is the following:

International Journal of Data Mining & Knowledge Management Process (IJDKP) Vol.6, No.2, March 2016```
SELECT some attributes
FROM G
WHERE selection condition
WITH data quality feature condition
ORDER BY F (P_1, .. P_m)
LIMIT k
```

A SQL example for Quantitate Single-Feature top-*k* selection query (Q4) is the following:

```
Select SName, SAddress, DOB, SupName, DName
From G
With AlternativeFactCompleteness ≥ 0.65
Order by AlternativeFactCompleteness desc
Limit 3
```

- **Qualitative Single-Feature Top-K Selection Query**

In Qualitative Single-Feature Top-K Selection Query, Data quality features values are represented as qualitative values.

A SQL example for Qualified Single-Feature top-k selection query (Q5) is the following:

```
Select SName, SAddress, DOB, SupName, DName
From G
With AlternativeFactCompleteness is high
Order by AlternativeFactCompleteness desc
Limit 3
```

**Received user message:**

High represents AlternativeFactCompleteness ≥ 0.65

Using the AlternativeAggregatedMetric table 26 that satisfies (Q4 and Q5), the ranked alternatives are presented in table 33

Table 33. Top 3 Alternatives Ranked according to FactCompleteness

| AlternativeName | Alternative FactCompleteness |
|---|---|
| Alternative2 | 0.95 |
| Alternative6 | 0.82 |
| Alternative3 | 0.68 |

**II. Multi- Feature Top-K Selection Query**

In this type, queries include many data quality features as a condition, so we will build AlternativeAggregatedMetric table from QueriedDataSourceAssessmentMetric table but after first and second pruning according to specified quality condition in user query and then return to user alternatives ranked according to user query case.

**Case1:** Data quality features specified in user's query are separated with (AND) and all are satisfied.





In this case, we will consider queried data source or alternative as qualified one if it satisfies all required data quality features together. The AlternativesAggregatedMetric table will be built from qualified alternatives and they will be ranked according to total score by TA algorithm.

**Case2:** Data quality features specified in user's query are separated with (AND) and the required level of quality for one or more of data quality features doesn't commensurate with the required level of quality for other data quality features or doesn't achieve.

In this case, there are no queried data sources or alternatives can return required query attributes with specified quality levels so, message will be sent to user to inform him that his required level of quality for query answering can't be satisfied with these data quality features together.

**Case3:** Data quality features specified in user's query are separated with (OR) and the required level of quality for all data features satisfied or the required level of quality for one or more of data quality features doesn't commensurate with the required level of quality for others data quality features or can't be achieved.

In this case, we will consider queried data source or alternative as qualified one if it satisfies at least one required data quality feature. The AlternativesAggregatedMetric table will be built from qualified alternatives and they will be ranked according to total score by TA algorithm.

This type of queries classified into two categories Quantitate Multi-Feature Top-K Selection Query and Qualified Multi-Feature Top-K Selection Query. We will introduce every category as following:

- **Quantitate Multi-Feature Top-K Selection Query**

In this type queries condition contains multi-features and their values are represented in quantitate way.

A SQL template for multi-Feature top-k selection query is the following:

```
SELECT some attributes
FROM G
WHERE selection condition
WITH data quality feature condition
ORDER BY F (P_1, . . P_m)
LIMIT k
```

A SQL example for Quantitate Multi-Feature top-k selection query (Q6) in (Case1) is the following:

```
Select SName, SAddress, DOB, SupName, DName
From G
With AlternativeFactCompleteness ≥ 0.65 and AlternativeValidity ≥ 0.65 and
AlternativeAccuracy ≥ 0.65
Order by AlternativeFactCompleteness desc, AlternativeValidity desc,
AlternativeAccuracy desc
Limit 3
```

- **Qualified Multi-Feature Top-K Selection Query**

In this type of queries, the query condition contains more than one feature but the values of data quality features are represented in qualitative way.

A SQL example for Qualified Multi-Feature top-k selection query (Q7) (Case1) is the following:





```
Select SName, SAddress, DOB, SupName, DName
From G
With AlternativeFactCompleteness high and AlternativeValidity high and
AlternativeAccuracy high
Order by AlternativeFactCompleteness desc, AlternativeValidity desc,
AlternativeAccuracy desc
Limit 3
```

**Received User message:**

High Validity represents AlternativeValidity ≥ 0.65, High Fact Completeness represents AlternativeFactCompleteness ≥ 0.65 and High Accuracy represents AlternativeAccuracy ≥ 0.65

To deal with Multi-Features Top-K Selection Queries, we should build Lists (tables); every list contains alternatives from alternative aggregated metric and it ranks descending according to one of data quality features' scores that is required in user query, so we need to combine these ranking to produce global ranking.[9]

In our work, we choose **Threshold Algorithm (TA)** proposed by Fagin et.al.2001 as ranking algorithm[10]. It considers famous, simple and elegant Top-K algorithm, it considers the basic algorithm for all next variants, it is applicable for queries where the scoring function is monotonic, it is based on an early-termination condition and it evaluates top-k queries without examining all the tuples. This algorithm is presented as following:

```
Algorithm: TA [Fagin et al. 2001]
(1) Do sorted access in parallel to each of the m sorted lists L_i. As a new object (o) is seen
under sorted access in some list, do random access to the other lists to find P_i (o) in every
other list L_i. Compute the score F (o) = F (P_1, . . . , P_m) of object o. If this score is among the
k highest scores seen so far, then remember object o and its score F (o).
(2) For each list L_i, let P_i be the score of the last object seen under sorted access. Define the
threshold value T to be F (P_1, . . . , P_m). As soon as at least k objects have been seen with
scores at least equal to T, halt.
(3) Let A_k be a set containing the k seen objects with the highest scores. The output is the
sorted set {(o, F (o)) |o ∈ A_k}.
```

Figure 3.Threshold algorithm (TA)[11]

Using the AlternativeAggregatedMetric table 26 that achieves (Q6 and Q7) and applying TA algorithm to following lists, the ranked alternatives will be presented in table 36

L1

| AlternativeName | AlternativeFactCompleteness |
|---|---|
| Alternative2 | 0.95 |
| Alternative6 | 0.82 |
| Alternative3 | 0.68 |

L2

| AlternativeName | AlternativeValidity |
|---|---|
| Alternative2 | 0.95 |
| Alternative6 | 0.79 |
| Alternative3 | 0.62 |

L3

| AlternativeName | AlternativeAccuracy |
|---|---|
| Alternative2 | 0.95 |
| Alternative6 | 0.79 |
| Alternative3 | 0.62 |

1) Sorted access in parallel to each of the 3 sorted lists**.**

L1

| AlternativeName | AlternativeFactCompleteness |
|---|---|
| Alternative2 | 0.95 |
| Alternative6 | 0.82 |
| Alternative3 | 0.68 |

L2

| AlternativeName | AlternativeValidity |
|---|---|
| Alternative2 | 0.95 |
| Alternative6 | 0.79 |
| Alternative3 | 0.62 |

L3

| AlternativeName | AlternativeAccuracy |
|---|---|
| Alternative2 | 0.95 |
| Alternative6 | 0.79 |
| Alternative3 | 0.62 |



International Journal of Data Mining & Knowledge Management Process (IJDKP) Vol.6, No.2, March 2016

2) For every new object o is seen under sorted access in some list, do random access to the other lists to find $P_i$ (o) in every other list $L_i$.

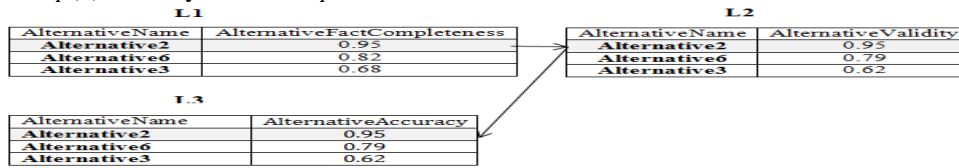

3) Compute the score F (o) = F ($P_1, \ldots, P_m$) of object o. If this score is among the *k* highest scores seen so far, then remember object o and its score F (o).
   Assume that ranking Function is sum, so F (Alternative 2) = 2.85

4) The threshold value T is F ($P_1, \ldots, P_m$) for the scores of the last seen object
   The threshold value T = 2.85
   **Because** F (Alternative2) = the threshold value T = 2.85 so, Alternative2 will put in $A_k$.

5) $A_k$ Is a set containing the *k* seen objects with the highest scores

Table 34. $A_k$ with k seen objects with the highest scores

| O | F(O) |
|---|---|
| Alternative2 | 2.85 |

Complete the same steps for all remaining objects.

6) 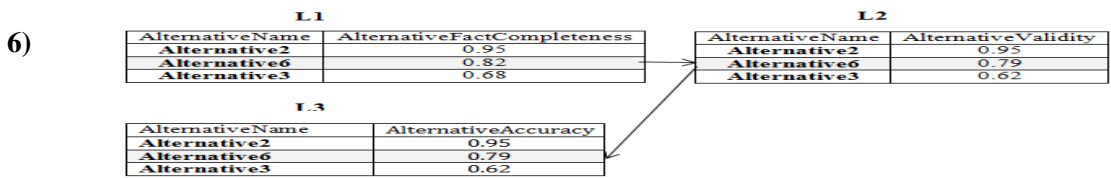

7) F (Alternative6) = 2.4
8) The threshold value T = 2.4
9) **Because** F (Alternative6) = the threshold value T = 2.4 so, Alternative6 will put in $A_k$ and rank the existing alternatives in descending order in

Table 35. $A_k$ with k seen objects with the highest scores

| O | F(O) |
|---|---|
| Alternative2 | 2.85 |
| Alternative6 | 2.4 |

10) 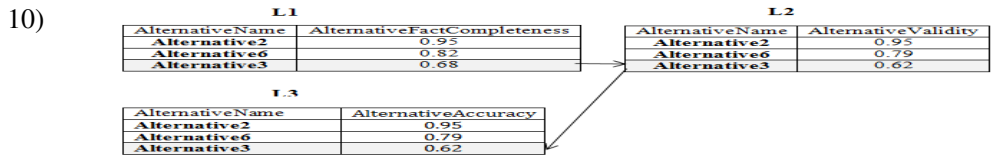

11) F (Alternative3) = 1.92
12) The threshold value T = 1.92
13) **Because** F (Alternative3) = the threshold value T = 1.92 so, Alternative3 will put in $A_k$.
    Alternative2, Alternative6 and Alternative3 will put in descending order in $A_k$

In our example, the alternatives ranking is the same for all data quality features that specified in user query so we can directly say that top-3 ranking alternatives that satisfy user requirements from quality are Alternative2, Alternative6 and Alternative3.

Table 36. $A_k$ with k seen objects with the highest scores

| O | F(O) |
|---|---|
| Alternative2 | 2.85 |
| Alternative6 | 2.4 |
| Alternative3 | 1.92 |




**Duplicate Detection and Data Fusion Process**

Some of Top-K Ranked Alternatives produced by DIRA consist of one qualified queried data source and others consist of more than one queried data source, those containing more than one queried data source will pass to duplicate detection and data fusion algorithms that will run on their results respectively then these results will be re-evaluated using assessment metrics that will use equations 5, 7, 12, 13 and 14 which are presented in section1 and section 3.2 to be added to alternatives with one qualified queried data source to re-rank to return final top-k alternatives.

## 4. CONCLUSION

In this paper, we presented data integration framework that integrates large number of available data sources with different levels of quality to return top-k qualified query answers from significant ones only.

This framework introduces new accurate type of completeness called fact-completeness that will be used in DIRA assessment module that works on four data quality features completeness, validity, accuracy and timeliness for early pruning of data sources under the required level of quality and retrieving data from only qualified ones, this framework also shortens processing time of duplicate detection and data fusion as they will work on only top-k alternatives with more than one queried data source not all available query results and it can be extended to include different types of data sources, add more data quality features and use different ranking algorithm.